\documentclass[submission,copyright,creativecommons]{eptcs}
\providecommand{\event}{CREST 2018} 
\usepackage{breakurl}             
\usepackage{underscore}           

\usepackage{ltl}
\usepackage{subfigure}

\usepackage[normalem]{ulem}

\usepackage{color}

\title{The Challenges in Specifying and Explaining Synthesized Implementations of Reactive Systems\thanks{This work was partly funded by the European Research Council (ERC) Grant OSARES (No. 683300) }}
\author{Hadas Kress-Gazit
\institute{Sibley School of Mechanical and Aerospace Engineering\\
Cornell University, Ithaca, NY, USA\\
}
\email{hadaskg@cornell.edu}
\and
Hazem Torfah\thanks{Corresponding author}
\institute{Reactive Systems Group\\
Saarland University, Saarbr\"ucken, Germany}
\email{torfah@react.uni-saarland.de}
}
\def\titlerunning{The Challenges in Reactive Synthesis}
\def\authorrunning{H. Kress-Gazit \& H. Torfah}
\begin{document}
\maketitle

\begin{abstract}
In formal synthesis of reactive systems an implementation of a system is automatically constructed from its formal specification. The great advantage of synthesis is that the resulting implementation is correct by construction; therefore there is no need for manual programming and tedious debugging tasks. 
Developers remain, nevertheless, hesitant to using automatic synthesis tools and still favor manually writing code. 
A common argument against synthesis is that the resulting implementation does not always give a clear picture on what decisions were made during the synthesis process. The outcome of synthesis tools is mostly unreadable and hinders the developer from understanding the functionality of the resulting implementation. 
Many attempts have been made in the last years to make the synthesis process more transparent to users. Either by structuring the outcome of synthesis tools or by providing additional automated support to help users with the specification process. 

In this paper we discuss the challenges in writing specifications for reactive systems and give a survey on what tools  have been developed to guide users in specifying reactive systems and understanding the outcome of synthesis tools.
\end{abstract}

\section{Introduction}
\emph{Synthesis} is a procedure in which an implementation of a system is automatically constructed from a logical specification. The resulting implementation is correct by construction and no further coding tasks are needed. Synthesis allows developers to focus on determining \emph{what} a system should do rather than \emph{how} it should do it. The task of the developer thus is shifted from writing a program that implements the system to writing a specification for it. This comes with the big advantage of allowing systems to be analyzed at  early design stages and disposes of tedious and costly  implementation efforts in later stages.

In the last decade, the theoretical ideas of synthesis have been translated into several
tools (cf. \cite{10.1007, conf/cav/FaymonvilleFT17, Unbeast10.1007, ACACIA, RATSY,DBLP:conf/cav/EhlersR16}). The tools have made it possible to tackle real-world design problems, such as the synthesis of an arbiter for the AMBA AHB bus, an open industrial standard for the on-chip communication and management of functional blocks in system-on-a-chip designs, or the IBM generalized buffer, which was synthesized from a specification written in PSL~\cite{Bloem07automatichardware}.   
Nevertheless, synthesis tools have barely been used outside the scientific community. 
Developers are hesitant to use automatic synthesis, and rather rely on self-created and self-maintained code, or use established legacy code.  
A common argument against synthesis is the high structural complexity of the resulting implementation. In most cases, synthesized implementations are not easy to follow and  do not allow  to structurally reason about the functionality of the system nor backtrack any mistakes introduced in its specification. The outcome of synthesis tools thus remains as a black box for developers that is hard to explore manually and where retracing relevant aspects of the input specification becomes infeasible.

One might argue that it is not the role of synthesis to provide understandable implementations more than correct ones. However, the correctness of synthesized implementation is only relative to the provided specification. In other words, the quality of the resulting implementation is only as good as its input specification. Understanding the functionality of the outcome  is thus vital for writing correct and high quality specifications. Tool support for refining specifications is thus vital for a correct synthesis outcome that indeed fulfills all the user's design intents.   

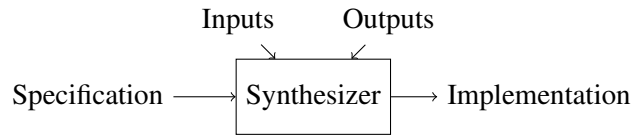
\begin{figure}[h]
\centering
\begin{tikzpicture}
	\node [draw, minimum height=1cm] (synthesis) at (0,0) {Synthesizer};
	\node (spec) at (-3,0) {Specification};
	\node (impl) at (3,0) {Implementation};
	\node (in) at(-1,1) {Inputs};
	\node (out) at(1,1) {Outputs};
	
	\path [->](spec) edge (synthesis);
	\path [->](synthesis) edge (impl);
	\path [->](in) edge (synthesis);
	\path [->](out) edge (synthesis);
\end{tikzpicture}
\label{fig:synthesis}
\caption{The Synthesis Problem}
\end{figure}

In this paper, we investigate the challenges in writing specifications for reactive systems and understanding automatically synthesized implementations. The setting we are interested in is given in Figure~\ref{fig:synthesis}, where given a set of inputs from the environment (sensors) and a set of outputs of the system (actuators), a synthesis tool constructs an implementation of the system that satisfies a high-level specification written over the inputs and outputs (correct reaction of actuators to sensor information).   
Specifications are usually given as formulas in a temporal logic that define the relations between inputs and outputs. Implementations are realizations of these relations represented as transducers (Mealy or Moore machines). 

Issues with the specifications in the synthesis process are captured by the problems of unrealizability and completeness of specifications. When writing specifications, one might over-specify  the systems, such that, no implementation can realize the system's specification. One might also under-specify parts of the system which results in synthesized implementations that satisfy the given specification but still do not meet the designers intents, i.e., they behave not as the designer expected  for certain input scenarios. Challenges on the implementation side involve the understandability of the resulting implementation and transparency regarding why the specific implementation was chosen from the set of all possible correct implementations.

We present a series of works that have addressed the construction of more structured and understandable implementations. 
In general, we can summarize the concerns using two questions:
(1)
 How do we assure that the synthesized implementation, which is one of many, is one that corresponds to the user's expectations? 
(2) How do we support the user in writing correct specifications, that include all the relevant aspects needed for the construction of  an implementation with all of the intended functionality?

We give an overview on challenges that we face on both the specification and implementation side of the synthesis process. We describe methods that are used for the analysis of specifications. Either by pointing out erroneous cores in a specification or indicating what assumptions have not been considered by the user. Tools can, for instance, return minimal specification revisions to make an unrealizable specification realizable. Dually, they should also identify vacuous parts of specifications, and help to eliminate ambiguities in the specification. Regarding the outcome of synthesis tools, we raise the question of how to determine the quality of a resulting implementation: Are there other artifacts that can be additionally generated 
to aid the user in understanding or validating the implementation and the specification?  Are there understandable witnesses that  validate the black-box implementation obtained as the output of the synthesis tool?
What further metrics can be used to debug specifications, reason about implementations and facilitate the composition of the implementation in a larger system?

\section{Writing Specifications for Reactive Systems}

Reactive systems are those systems that react to inputs from an  environment. A specification of a reactive system thus defines how an implementation of the system should behave in response to inputs  from this environment.   A specification usually includes assumptions on the environment, which define the scope in which the implementation should behave correctly, and guarantees that define the correct behavior of the system under those assumptions. A synthesis procedure tries then to construct an implementation that fulfills all the guarantees under the assumptions declared over the environment. 

 In general, two types of problems may occur when specifying a reactive system. One might over-specify the system making the specification become unrealizable, i.e., there is no implementation that satisfies the specification.
 One might under-specify the system by leaving out many relevant details, that are crucial for a synthesized implementation that behaves as the user expected. 
 
 The first type of error is detected by the synthesis tools. If a specification is unrealizable, the synthesizer is not able to return an implementation and may return a counterexample  that captures the change in the inputs that will make the system fail. The second type of error is harder to  detect because the synthesizer terminates with an implementation of the system, but gives no further information about how the unspecified parts of the implementation were constructed. 

In the following we discuss both types of specification errors and show how synthesis tools can potentially leverage each of these errors for the purpose of correcting specifications.  

\subsection{Unrealizability of Specifications} 
A specification is unrealizable, either because it is unsatisfiable, i.e., the value of the specification is equal to false due to inconsistencies in the specification, or  
it is unrealizable because there exists no implementation that behaves correctly over all input sequences of the environment.
Consider for example an arbiter over two processes as  given in Figure~\ref{fig:arbiter}, and consider a specification for the arbiter given by the conjunction of the LTL formulas \textit{response} and \textit{access}:
\begin{figure}[h]
\centering
\begin{tikzpicture}
	\node [draw, rounded corners=.09cm] (ctrl) at (0,0) {arbiter};
	\node [draw, fill=black] (p1) at (-1,-1.5) {{\color{white}$p_1$}};
	\node [draw, fill=black] (p2) at (1,-1.5) {{\color{white}$p_2$}};
	\path [->] (p1) edge [bend right =15] (ctrl) node [right =.7cm, below =-.6cm]{$r_1$};
	\path [->] (ctrl) edge [bend right =15] (p1) node [left = .8cm, above=-.7cm] {$g_1$};
	\path [->] (p2) edge [bend left =15] (ctrl) node [left =.7cm, below =-.6cm]{$r_2$};
	\path [->] (ctrl) edge [bend left =15] (p2) node [right = .9cm, above=-.7cm] {$g_2$};	
	\node (response) at (5,0){\textit{response}:~$\bigwedge \limits_{i \in \{1,2\}}\LTLsquare (r_i\rightarrow \LTLdiamond g_i)$ };
	\node (access) at (5,-1){\textit{access}:~$ \LTLsquare(\neg a \rightarrow \neg g_1 \wedge \neg g_2)$};
	\node (w) at (-1.5,0) {$a$};
	\path [->] (w) edge (ctrl) {};
\end{tikzpicture}
\caption{An arbiter over two processes and a specification for the arbiter given as a conjunction of the formulas \textit{response} and \textit{access}.}
\label{fig:arbiter}
\end{figure}
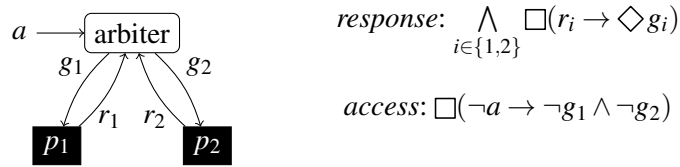

\noindent The arbiter controls the access of the processes $p_1$ and $p_2$ into a shared resource. The processes $p_1$ and $p_2$ request access to the resource via the signals $r_1$ and $r_2$, respectively. The arbiter grants access to the processes using the respective signals $g_1$ and $g_2$. A further external signal $a$ determines when the resource can be accessed. The signals $r_1,r_2$ and $a$ compose the inputs of the environment and $g_1$ and $g_2$ are the outputs of the system. 

The specification $\texttt{response} \wedge \texttt{access} $ is unrealizable. The environment can always set the input signal $a$ to false, forbidding the arbiter from sending any grants $g_1$ or $g_2$. Thus, there is no implementation that satisfies the specification $\textit{response}$ for an input of the environment, where a request $r_i$ has been sent to the arbiter by one of the processes and where the signal $a$ is always false\footnote{The specification is an example of an unrealizable nevertheless satisfiable specification. 
An example input sequence that has a corresponding satisfying output sequence is for example the sequence $\{ a ,g_1, r_1\}^\omega$.}.   

 In this section we present a list of artifacts for explaining unrealizability, methods for detecting unrealizable cores of specifications, and how to modify unrealizable specifications to get realizable ones.

\subsubsection{Detecting and  mitigating unrealizability}
Checking the realizability of specifications can be seen as a game theoretic problem where two players, the environment and the system, interchangeably produce input and outputs over an infinite duration~\cite{Buchi1990}. Without loss of generality we assume that in our setting the system player starts the game, by initializing the values of the atomic propositions of the system. An implementation of a system for a given specification is a winning strategy for the system player in that game. A specification is realizable if there exists a winning strategy for the system. A specification is unrealizable if for each strategy of the system, there is an input sequence of the environment where the strategy loses the game, i.e., for which the strategy is forced to produce an output that violates the specification. Consider the unrealizable specification given in Figure~\ref{fig:arbiter}. No matter what strategy the systems chooses, the environment challenges the strategy with the input sequence $\{r_1\}\{\}^\omega$, where the process $p_1$ sends a request to the arbiter but where the signal $a$ is always set to false. As any correct strategy must give a grant, but at the same time is not allowed to, because the signal $a$ is always false, no strategy is able to fulfill both the specifications $\texttt{response}$ and $\texttt{access}$. 

If a specification is unrealizable, then there is a set of input sequences for which no matter what strategy the system chooses, the strategy will produce a violating
 output sequence on at least one of those input sequences. We call such a set of input sequences a counterexample set for the unrealizable specification. 
In the setting considered in this paper, finding a counterexample set can be done by solving the synthesis game. For more general settings such as distributed architectures or settings with incomplete information the problem is in general undecidable \cite{hardtosynthesize}.

The counterexample set can grow infinitely large. 
 Consider for example the following architecture and LTL specification:
 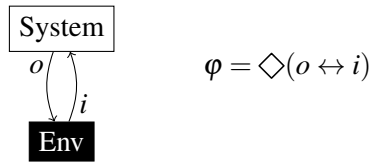
\begin{figure}[h]
 \centering
 \begin{tikzpicture}
 	\node [draw] (sys) at (0,0) {System};
 	\node [draw, fill=black] (env) at (0,-1.5) {{\color{white}Env}}; 
 	\path [->]  (sys) edge [bend right =20] (env) node [below=.5cm, left=0.1cm ] {$o$};
 	\path [->]  (env) edge [bend right =20] (sys) node [above=0.5cm, right=0.1cm] {$i$};
 	\node at (3,-.5) {$\varphi = \LTLeventually (o \leftrightarrow i)$};
 \end{tikzpicture}
 \caption{An unrealizable LTL with an infinite set of counterexamples}
 \label{fig:infunrealizable}
 \end{figure} 
 
 \noindent The specification requires the environment to send an input $i$ if and only if the system outputs $o$. This specification is unrealizable as the system has no control over the environment\footnote{Remember the system moves first.}. 
 A counterexample set for the specification is given by the set $\Gamma= (2^{\{i\}})^\omega$, and there is no finite set $\Gamma'\subset \Gamma$ that is a counterexample set for $\varphi$. Assume there is a finite set $\Gamma'$ that is a counterexample set for $\varphi$. Because,  $\Gamma'$ is finite, there is a position $j$ such that all prefixes of length $j$ of the sequences in $\Gamma'$ are pairwise different.  As the sequences in $\Gamma'$ can be distinguished at position $j$ we can choose a strategy for the system that assigns the value of $o$  at position $j+1$ to  true if the input $i$ at this position is true, and otherwise sets $o$ to false. This strategy satisfies the property $\varphi$ over all sequences in $\Gamma'$ and thus $\Gamma'$ cannot be a counterexample set for $\varphi$ \cite{FT15}. 
 
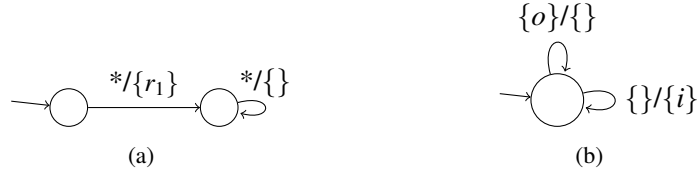
\begin{figure}
 \centering
 \subfigure[]{
 	\begin{tikzpicture}
 	    \node (init) at (-0.9,0.1) {};
 		\node [circle, draw, minimum size = 0.5cm](q0) at (0,0) {};
 		\node [circle, draw, minimum size = 0.5cm](q1) at (2,0) {};
 		\path [->] (init) edge (q0);
 		\path [->] (q0) edge [] node [above]{*/\small $\{r_1\}$} (q1);
 		\path [->] (q1) edge [loop right] node [right, above] {*/$\{\}$}(q1);
 	\end{tikzpicture}
 } 
 \hspace{2cm}
 \subfigure[]{
 	\begin{tikzpicture}
 		\node (init) at (-0.9,0.1) {};
 		\node [draw, circle, minimum size = 0.7cm](ni) at (0,0){};

 		\path [->] (init) edge (ni);
 		\path [->] (ni) edge [loop above] node [left, above] {$\{o\}$/$\{\}$}(ni);
 		\path [->] (ni) edge [loop right] node [right] {$\{\}$/$\{i\}$}(ni);
 	\end{tikzpicture}
 }
 \caption{Counterstrategies for the unrealizable specifications $\texttt{access}\wedge \texttt{response}$ in Figure~\ref{fig:arbiter} and~$\varphi$~in~Figure~\ref{fig:infunrealizable} given as Mealy machines.}	
 \label{fig:cex}
 \end{figure}

 A convenient finite representation of the possibly infinite set of counterexamples can be given by a \emph{counterstrategy}. A counterstrategy is a winning strategy for the environment, and it is computed by solving the synthesis game for the environment player instead of the system player.   A counterstrategy for the unrealizable specification in Figure~\ref{fig:arbiter} is given in Figure~\ref{fig:cex}(a). The strategy responds to the first output of the system by assigning the input $r_1$ to true and assigns all subsequent inputs to false independent of the chosen outputs by the system. A counterstrategy for the specification $\varphi$ in Figure~\ref{fig:infunrealizable} is given in Figure~\ref{fig:cex}(b). The strategy assigns the input signal $i$ to true if the system outputs false and to true otherwise. In this way the system will never fulfill the specification $\varphi$. 

Complex specifications may lead to large and complex counterstrategies that are difficult to inspect manually. In many cases, there is no need to consider the whole counterstrategy to infer what parts of the specification are unrealizable.  A smaller set of input sequences might already suffice to decide the unrealizability of the specification. Some techniques rely on pruning parts of the counterstrategy that are irrelevant for its unrealizability in order to make the counterstrategy more readable \cite{Brazdil2015}. Further works suggested to only return a sufficient set of input scenarios of the environment instead of returning the whole counterstrategy. An alternative, for example, are countertraces  \cite{5351127}, 
which are fixed input traces for which there is no output trace fulfilling the specification. One problem with countertraces nevertheless is that they are hard to compute and sometimes they do not exist. 
In case of safety properties one can compute a finite counterexample set of finite sequences using the symbolic method presented in \cite{FT15}. The finite sequences resemble scenarios where the system violates the safety property after  finitely many steps.  The method involves a procedure that incrementally increases the bound on the size of input sequences until a counterexample set is found. The big advantage of this  method is that it also provides a semi-decision procedure for the unrealizability problem over undecidable distributed architectures.

 Treating unrealizability can also be done by directly analyzing the specification itself, for example by identifying  unrealizable cores of the specification (e.g. ~\cite{SCHUPPAN2012908,Koenighofer2011,cimatti,DBLP:journals/arobots/LignosRFMK15} ).
 An unrealizable core is a sub-specification that is unrealizable on its own. Consider our arbiter example again and let:
 \begin{align*}
 	\varphi_1 & =  \LTLsquare (r_1 \rightarrow \LTLdiamond g_1) \\
 	\varphi_2 & =  \LTLsquare (r_2 \rightarrow \LTLdiamond g_2) \\
 	\varphi_3 & =  \LTLsquare (\neg a \rightarrow \neg g_1 \wedge \neg g_2)
 \end{align*}
\noindent The specification contains the following minimal unrealizable cores: $C_1 =\{\varphi_1,\varphi_3\}$ and $C_2=\{\varphi_2,\varphi_3\} $. To make the specification realizable, one has to resolve both the conflicts $C_1$ and $C_2$. This can be done by either weakening the specifications $\varphi_1$ and $\varphi_2$, for example, by relaxing the eventuality to $\LTLsquare (r_1 \rightarrow (\neg a \vee g_2)  W g_1 )$ and  $\LTLsquare (r_2 \rightarrow (\neg a \vee g_1)  W g_2 )$ using the \emph{weak until} operator $W$. In this way, the requests $r_1$ and $r_2$ must be answered by the respective grants, as soon as the access signal  $a$ becomes true, otherwise the specification specifically states that no grants are to be given. 
Another possibility to make the specification realizable is by restricting the behavior of the environment. The main reason why the specification is not realizable is because the environment can choose not to set the signal $a$ to true. However, this assumption on the behavior of the environment is not necessarily realistic. We can add a further assumption that states that the environment will grant access to the shared resource an infinite number of times, namely  $\varphi_4 = \LTLsquare \LTLdiamond a$, making the specification realizable.

Detecting unrealizability is only the first step. As important is assisting the developer in repairing the specification. A series of works \cite{PatternBasedRefinement,AssumeGuarantee, 5970509,alurmoarref, gist, cheng2014g4ltl,6301745,chaterjee2008, fukaya} introduced frameworks that leverage the artifacts above  to turn an unrealizable specification into a realizable one.

As a specification for a reactive systems includes assumptions on the environment and guarantees to be fulfilled by the system, making a specification realizable can be done by either strengthening the assumptions on the environment or weakening the guarantees of the system. 
Strengthening the assumptions on the environment is done by adding further assumptions that remove certain scenarios for which the specification is unrealizable. 
Weakening the guarantees is done by tolerating additional behaviors of the system.  Most approaches rely on a counterexample-guided refinement loop to learn the new assumptions \cite{PatternBasedRefinement,AssumeGuarantee, 5970509,alurmoarref, gist, cheng2014g4ltl}. In each refinement loop a  counterstrategy is used to extract new assumptions for the environment. Some  approaches try to directly learn assumptions on the environment by first computing a safety assumption that removes a minimal set of environment edges from the game graph, and then computing a liveness assumption that puts fairness conditions on some of the remaining environment edges~\cite{chaterjee2008, fukaya}.

 An interactive approach to identifying the cause of failure in an unrealizable specification was presented in \cite{6301745}. Here, a game-based approach is presented where the user attempts to fulfill a robot specification against an adversarial environment. The idea of the approach is to highlight bad portions of the specification and identify 
example executions for the environment that make the system fail.

\subsection{Incomplete Specifications} A common error when specifying systems is to under-specify them. In this case, synthesis tools will return an implementation for the given specification but that may still behave different than the user expected. Revisiting the two process arbiter given in Figure~\ref{fig:arbiter}, assume we want to synthesize an implementation for the arbiter that is mutually exclusive and where every request is guaranteed to be answered eventually. A specification for the arbiter can be given by the respective LTL formulas $\LTLsquare(\neg g_1 \vee \neg g_2)$, and $\LTLsquare (r_1 \rightarrow \LTLdiamond g_1)$ and $\LTLsquare (r_2 \rightarrow \LTLeventually g_2)$. A possible outcome of the synthesis tool could be an implementation as given in Figure~\ref{fig:arbiterImpl}(a). The implementation returns immediately a grant $g_1$ every time there is a request $r_1$ and a grant $g_2$ whenever there is a request $r_2$. If both request $r_1$ and $r_2$ occur at the same time, the implementation prioritizes process $p_1$ by first giving a grant $g_1$ and as soon as process $p_1$ is done, it grants $p_2$ access to the shared resource. The decision to give $p_1$ priority was made by the synthesis tool.  If the user is not happy with prioritizing process $p_1$ then an additional specification must be added by the user to handle simultaneous requests more adequately. 

The implementation in Figure~\ref{fig:arbiterImpl}(a) is not the only realization of the arbiter's specification. Figure~\ref{fig:arbiterImpl}(b) shows another implementation for the arbiter that interchangeably returns grant $g_1$ and $g_2$ without considering what requests were made by the processes. This means that the grants are given even if no requests were made by the processes, which is not necessarily what the user intended by the specification. 
This further means that the specification was not explicit enough on whether a grant depends on the requests, as  
in the previous implementation. To avoid the construction of such implementations, the specification must be refined. \\

One possible modification could be to change the specifications describing the responsiveness of the arbiter to   $ \LTLsquare (r_1 \leftrightarrow \LTLdiamond g_1)$ and $ \LTLsquare (r_2 \leftrightarrow \LTLeventually g_2)$. In this way it is more likely that an implementation such as the one in Figure~\ref{fig:arbiterImpl}(a)is enforced. \\

\begin{figure}[t]
\centering
\subfigure[]{
	\begin{tikzpicture}[scale=0.9]
		\node [] (init) at (-1.9,0.3){}; 
		\node [draw, circle, minimum size =1.1cm](0) at (-1,-0.4) {\small\{\}};
		\node [draw, circle, minimum size =0.8cm](1) at (1.3,1.2) {\small$\{g_1\}$};
		\node [draw, circle, minimum size =0.8cm](2) at (2,-2.0) {\small$\{g_2\}$};
		\node [draw, circle, minimum size =0.8cm](3) at (4.4,1.3) {\small$\{g_1\}$};
		\node [draw, circle, minimum size =0.8cm](4) at (6.9,-1) {\small$\{g_2\}$};
		
		\path [->] (init) edge (0){};
		\path [->] (0) edge [loop left] node [left]{\tiny$\{\}$}(0);
		\path [->] (0) edge [bend right =10] node [right=7, below] {\tiny$\{r_1\}$} (1);
		\path [->] (0) edge [bend right =60] node [left=7, below] {\tiny $\{r_2\}$} (2);
		\path [->] (0) edge [bend left =100] node [right=13,above=6] {\tiny$\{r_1,r_2\}$} (3);

		\path [->] (1) edge [loop above] node [left=17, below=2]{\tiny$\{r_1\}$}(1);
		\path [->] (1) edge [bend right =10] node [left=7, above] {\tiny$\{\}$} (0);
		\path [->] (1) edge [bend right =10] node [left=9,below] {\tiny$\{r_2\}$} (2);
		\path [->] (1) edge [bend left =10] node [left=7, above] {\tiny $\{r_1,r_2\}$} (3);
		
		\path [->] (2) edge [loop below] node [below]{\tiny$\{r_2\}$}(2);
		\path [->] (2) edge [bend left =20] node [right=10, below=4] {\tiny$\{\}$} (0);
		\path [->] (2) edge [bend right =10] node [right=15, below] {\tiny$\{r_1\}$} (1);
		\path [->] (2) edge [bend left =10] node [left=10, above=8] {\tiny$\{r_1,r_2\}$} (3);
		
		\path [->] (3) edge [bend left =50] node [left=14,below=7] {\tiny $\{r_1\},\{r_1,r_2\}$} (4);
		\path [->] (3) edge [bend left =10] node [below=20] {\tiny $\{r_2\}, \{\}$} (2);
		
		\path [->] (4) edge [bend right =90] node [above, right=7] {\tiny$\{\},\{r_1\}$} (1);
		\path [->] (4) edge [bend left =40] node [left=5, below=8] {\tiny$\{r_2\},\{r_1,r_2\}$} (3);
	\end{tikzpicture}
}
\hspace{1.3cm}
\subfigure[]{
	\begin{tikzpicture}
		\node [] (init)at (-1,0.3){}; 
		\node [draw, circle, minimum size =0.9cm](1) at (0,0) {\small$\{g_1\}$};
		\node [draw, circle, minimum size =0.9cm](2) at (2,0) {\small$\{g_2\}$};
		\node [] at (0,-2.5) {};
		
		\path [->] (init) edge (1){}; 
		\path [->] (1) edge [bend left=20] node [above] {*}(2);
		\path [->] (2) edge [bend left=20] node [below] {*}(1);
		
	\end{tikzpicture}
}
\caption[]{Two different implementations for the specification $\LTLsquare(\neg g_1 \vee \neg g_2)\wedge \LTLsquare (r_1 \rightarrow \LTLdiamond g_1) \wedge \LTLsquare (r_2 \rightarrow \LTLeventually g_2)$.}
\label{fig:arbiterImpl}
\end{figure}

Completeness of specifications cannot be defined formally, as it is dependent on the user's design intents. Nevertheless, with respect to this, we can say that a specification is complete if no implementation that satisfies the specification is incorrect with respect to the user's intent. In general, debugging an incomplete specification is a multistage refinement process. In the following we present some methods on how to aid the user throughout this process to construct a complete specification.

\subsubsection{Detecting vacuity in specifications}
Different synthesis procedures result in different implementations for the same specification. The reason for that is that parts of the implementation that are not explicitly defined by the specification are completed by the underlying decision procedure of the synthesis tools. For example, in the implementation in Figure~\ref{fig:arbiterImpl}(a), the synthesis procedure decided to set the values of the atomic propositions $g_1$ and $g_2$ to false in the initial state, as the specification did not explicitly state what the values of these atomic propositions should be. Another synthesis procedure could have chosen different values as long as mutual exclusion is ensured. 

 To understand which parts of the implementation were forced by the specification and which parts were decided by the synthesis procedure, one has to perform a coverage analysis on the resulting implementation. Intuitively, an atomic proposition of a state in a transition system is covered by the specification if changing the value of the atomic proposition in that state falsifies the specification \cite{coverage}. For example changing the value of the atomic proposition $g_1$ in the initial state of the transition system in Figure~\ref{fig:arbiterImpl}(a) from false to true  does not falsify the specification. Thus, the value of $g_1$ in the initial state is not covered by the specification. In the transition system in Figure~\ref{fig:arbiterImpl}(b) on the other hand, changing the value of $g_1$ does violate the specification.   
 
 Definitions of coverage range from qualitative definitions like the above to quantitative versions based on certain metrics \cite{Beer2001,Ben-David2015, Hoskote:1999:CES:309847.309936, coverage, Chockler:2008:CSS:1352582.1352588}.  A variant of coverage is one based on causality.  In the implementation in Figure~\ref{fig:arbiterImpl}(b) choosing $g_1$ to be true in the initial state forced $g_2$ to be true in the other state. Thus, the decision made in the other state is caused by the decisions made in the initial state. If changing the value of an atomic proposition $a$ in one state $q$ does not falsify the specification, one should check whether there is a set of states $Y$, such that, changing the value of $a$ and the values of atomic propositions in $Y$ falsifies the specification. If this is the case, then choosing the current values of the atomic propositions in $Y$ has a causal relation to choosing the value of $a$ in $q$.

Using the various coverage definitions the designer can examine synthesized implementations and modify the specification accordingly. This requires   several synthesis and refinement steps until a complete specifications is reached that enforces a desired implementation for the system, for example, to get an implementation as in Figure~\ref{fig:arbiterImpl}(a) instead of another implementation like in Figure~\ref{fig:arbiterImpl}(b). By taking a closer look into the arbiter's specification and the usual mechanism of requests and grants, it is  clear to a human observer that the user intended grants to be given upon request from the processes. A system that receives no requests from the processes should not send out unnecessary permissions to enter the shared resource. A smart synthesis algorithm  will construct an implementation that considers each part of the specification entered by the user and avoids implementations like the one in Figure~\ref{fig:arbiterImpl}(b), which vacuously satisfy the specification by ignoring parts of the specification, in this case  the values of the signals $r_1$ and $r_2$. We say that an implementation non-vacuously satisfies a specification if it satisfies the specification but not any strengthening of the specification \cite{DBLP:conf/vmcai/BloemCES17}. Instead of synthesizing any transition system that satisfies the specification, a good synthesis procedure would construct a non-vacuous implementation that covers all parts of the specification \cite{DBLP:conf/vmcai/BloemCES17}.

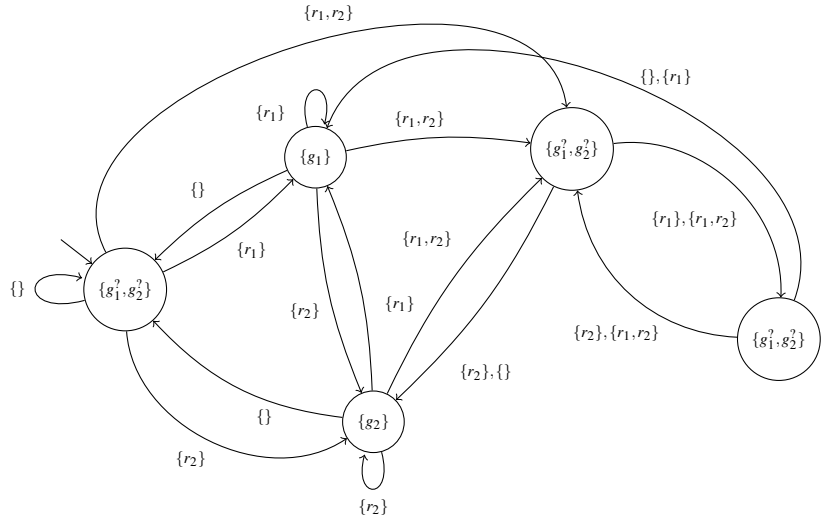
\begin{figure}[h]
\centering
\begin{tikzpicture}[scale=1.1]
		\node [] (init) at (-1.9,0.3){}; 
		\node [draw, circle, minimum size =1.1cm](0) at (-1,-0.4) {\tiny$\{g_1^? , g_2^?\}$};
		\node [draw, circle, minimum size =0.8cm](1) at (1.3,1.2) {\tiny$\{g_1\}$};
		\node [draw, circle, minimum size =0.8cm](2) at (2,-2.0) {\tiny$\{g_2\}$};
		\node [draw, circle, minimum size =0.8cm](3) at (4.4,1.3) {\tiny$\{g_1^? , g_2^?\}$};
		\node [draw, circle, minimum size =0.8cm](4) at (6.9,-1) {\tiny$\{g_1^? , g_2^?\}$};
		
		\path [->] (init) edge (0){};
		\path [->] (0) edge [loop left] node [left]{\tiny$\{\}$}(0);
		\path [->] (0) edge [bend right =10] node [right=7, below] {\tiny$\{r_1\}$} (1);
		\path [->] (0) edge [bend right =60] node [left=7, below] {\tiny $\{r_2\}$} (2);
		\path [->] (0) edge [bend left =100] node [right=13,above=6] {\tiny$\{r_1,r_2\}$} (3);

		\path [->] (1) edge [loop above] node [left=17, below=2]{\tiny$\{r_1\}$}(1);
		\path [->] (1) edge [bend right =10] node [left=7, above] {\tiny$\{\}$} (0);
		\path [->] (1) edge [bend right =10] node [left=9,below] {\tiny$\{r_2\}$} (2);
		\path [->] (1) edge [bend left =10] node [left=7, above] {\tiny $\{r_1,r_2\}$} (3);
		
		\path [->] (2) edge [loop below] node [below]{\tiny$\{r_2\}$}(2);
		\path [->] (2) edge [bend left =20] node [right=10, below=4] {\tiny$\{\}$} (0);
		\path [->] (2) edge [bend right =10] node [right=15, below] {\tiny$\{r_1\}$} (1);
		\path [->] (2) edge [bend left =10] node [left=10, above=8] {\tiny$\{r_1,r_2\}$} (3);
		
		\path [->] (3) edge [bend left =50] node [left=14,below=7] {\tiny $\{r_1\},\{r_1,r_2\}$} (4);
		\path [->] (3) edge [bend left =10] node [below=20] {\tiny $\{r_2\}, \{\}$} (2);
		
		\path [->] (4) edge [bend right =90] node [above, right=7] {\tiny$\{\},\{r_1\}$} (1);
		\path [->] (4) edge [bend left =40] node [left=5, below=8] {\tiny$\{r_2\},\{r_1,r_2\}$} (3);
	\end{tikzpicture}
\caption[]{A skeleton for the implementation in Figure~\ref{fig:arbiterImpl}(a) and the specification $\LTLsquare(\neg g_1 \vee \neg g_2)\wedge \LTLsquare (r_1 \rightarrow \LTLdiamond g_1) \wedge \LTLsquare (r_2 \rightarrow \LTLeventually g_2)$.}
\label{fig:skeleton}
\end{figure}

In general it is useful to inform the user on the decisions made during the synthesis process. This helps understand which parts were implemented independently by the synthesis procedure and which parts were forced by the specification. Synthesis tools thus need to provide  additional relevant information accompanied with each synthesized implementation. A first step towards this direction is the construction of skeletons for specifications \cite{Finkbeiner2016}. Skeletons are transition systems, where states are labeled with a three-valued assignment to the output variable: in each state an output can be true, false, or open, which means that the specification allows implementations with either value for a variable in that state. States with open variables indicate that additional constraints may be added to complete the specification according the user's intent. Skeletons can additionally be constructed with each synthesized implementation. For example, a skeleton for the transition system in Figure~\ref{fig:arbiterImpl}(a) is given in Figure~\ref{fig:skeleton}. Notice that from the skeleton we can read that the  implementation of the initial state and the decision to prioritize process $p_1$ are marked with "?", indicating that these choices were made by the synthesis procedure and were not explicitly determined by the specification. 

\subsubsection{Monitoring the implementation}
In many cases the environment assumptions may not be known to the user in full, which results in implementations with incorrect behavior. Many violations of the environment assumptions can be detected during runtime or during simulation. 
To better understand the violations, one can deploy monitors that give feedback on what caused the violation of these assumptions and modify the specification of the system accordingly.
In an automated feedback-based process, the specification of the system is augmented with new environment assumptions that are computed at runtime. Whenever the automated process fails, feedback is provided to the user, who is then asked to resolve the conflict by modifying the specification~\cite{DBLP:conf/rss/WongEK14,7139021}.

\section{Analyzing the Outcome of Synthesis Tools}
\begin{figure}[h]
\centering
\subfigure[][Manually written code]{
\includegraphics[width=0.16\textwidth]{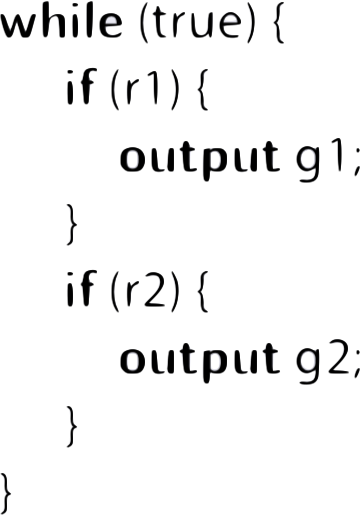}}
\hspace{2cm}
\subfigure[][Synthesized code by Acacia+]{
\includegraphics[width=0.36\textwidth]{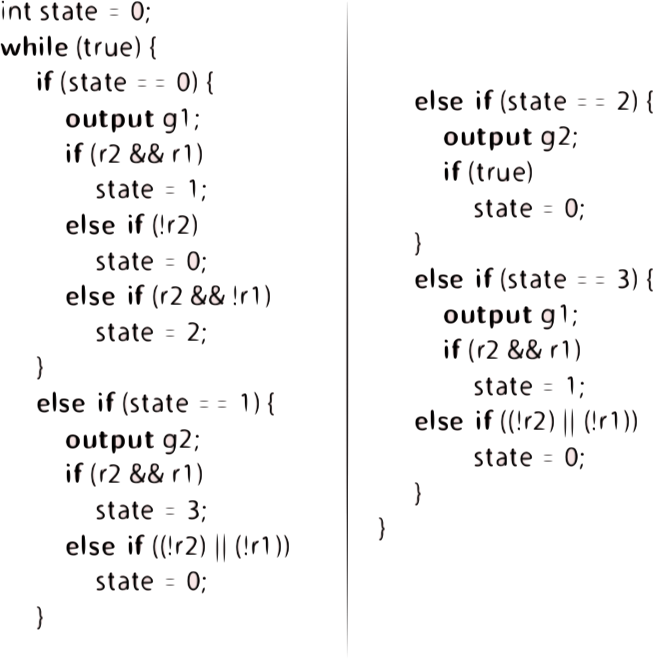}}
\caption{A manually written vs. a synthesized program for a two client arbiter  }
\label{fig:largeoutcome2}	
\end{figure}

\begin{figure}[h!]
\centering
\subfigure[][Manually written code]{
\includegraphics[width=0.16\textwidth]{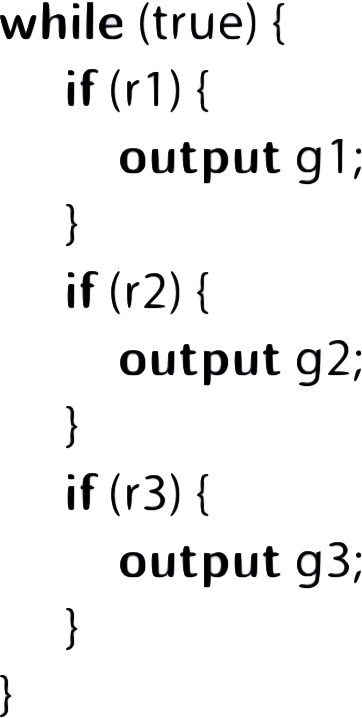}}
\hspace{2cm}
\subfigure[][Synthesized code by Acacia+]{
\includegraphics[width=0.40\textwidth]{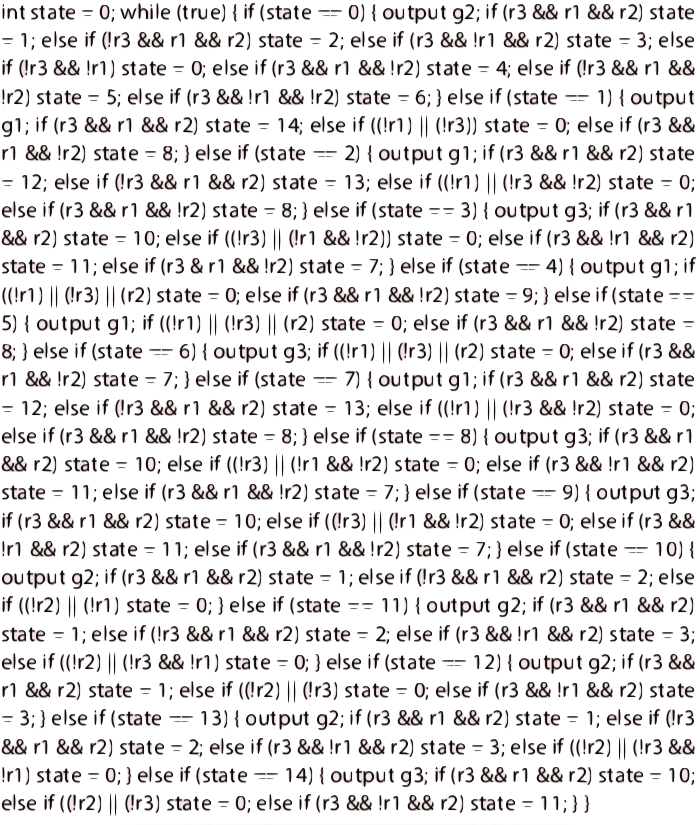}}
\caption{A manually written vs. a synthesized program for a three client arbiter }
\label{fig:largeoutcome3}	
\end{figure}

In most cases, the structure of an implementation produced by a synthesis tool is very complex and  hard to examine, and thus it is a challenge to convince the user that a synthesized implementation indeed does what it is actually supposed to do by just looking at it. 
Figures~\ref{fig:largeoutcome2} and \ref{fig:largeoutcome3} show  examples of synthesized and manually written implementation of two and three client arbiters. Notice that increasing the number of clients by one results in a large blow up in the synthesized implementation.

In order to make an implementation more understandable, synthesis tools must either synthesize structurally less complex implementations or provide the user with additional information that make the resulting implementation easy to follow. 
In the following we show some of the improvements that have been made to make the outcome of synthesis tools more  understandable for the developer.

\subsection{Representation of Implementations}
   
A synthesized implementation of a system from its specification is given by a transducer (a Mealy or Moore machine). Due to its large state space, there is a general trend to represent transducers  succinctly by binary decision diagrams (BDD) or circuits ~\cite{SYNTCOMP2017}. Such artifacts give  symbolic  representations of transducers that are easy to process but have the drawback of not mirroring the original functional choices of an implementation. Looking at a binary decision diagram, the developer will not be able understand the functionality of the implementation easily.
Many works have been devoted to minimizing or simplifying BDDs \cite{Akers:1978:BDD:1310167.1310815, GeneticAlgorithmsbdd, 206358}, but such operation are however notoriously difficult. Some also tried to use similar but more structured versions of BDDs to make the representation more explanatory \cite{tacasjantomas}. However, the structure remains too complex to explore manually. 

\begin{figure}
\centering
	\includegraphics[width=0.9\textwidth]{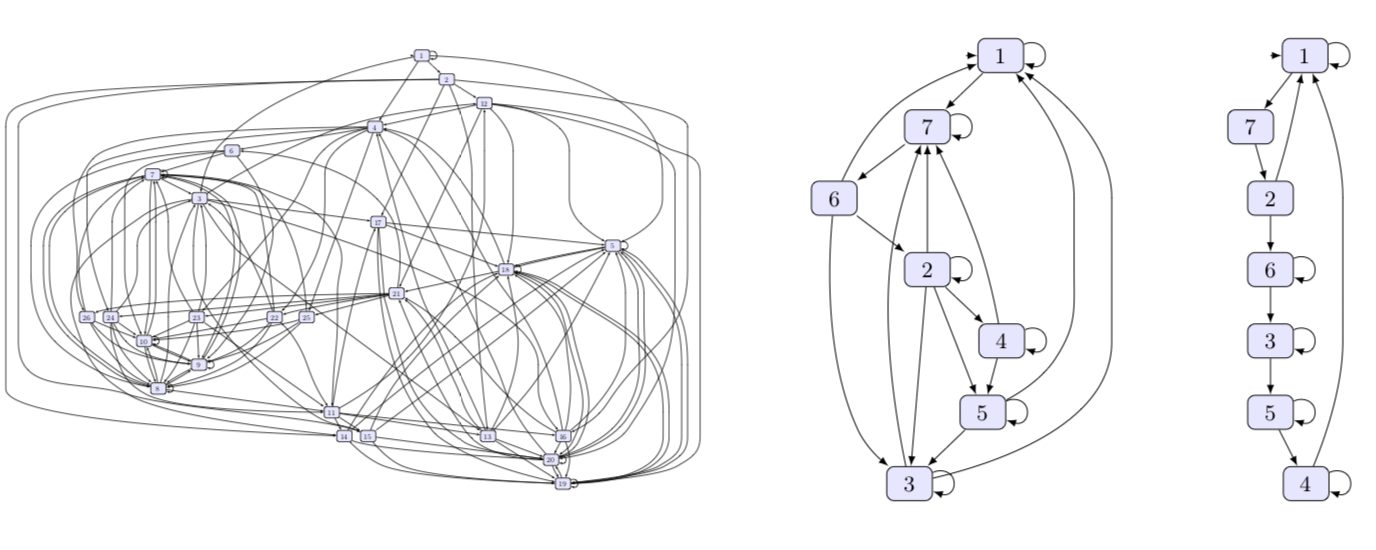}
\caption{\cite{FinkbeinerKlein16} Three implementations of the TBURST4 component of the AMBA bus controller. Standard synthesis with Acacia+ \cite{ACACIA} produces the state graph on the left with 14 states and 61 cycles. Bounded synthesis produces the graph in the middle with 7 states and 19 cycles. Bounded cycle synthesis, has 7 states and 7 cycles, which is the minimum.}
\label{fig:outputsensitivealg}
\end{figure}

 In general, the desire is not only to construct small but also  structurally simple and understandable implementations. 
To achieve this goal, algorithms are needed, which perform optimally not only in the input specification, but also in the structural complexity of the implementation, so called output-sensitive algorithms \cite{F16}. The first output-sensitive reactive synthesis algorithm was bounded synthesis \cite{FS13}. In bounded synthesis, the number of states of the implementation to be synthesized is an additional parameter to the synthesis algorithm. Minimal solutions are thus ensured by synthesizing implementations for incrementally increasing bounds.  Further metrics that help reduce the structural complexity of the implementations were introduced in \cite{FinkbeinerKlein16}. In addition to the size, the number of cycles in the state graph of the transducer is limited by a given bound. Reducing the number of cycles makes an implementation much easier to understand.
Figure~\ref{fig:outputsensitivealg} shows the different structural complexities of transducers synthesized using the bounded size, bounded cycle and a non-output sensitive algorithm.

Other works have tended to reduce the complex synthesis result to a much more understandable version by approximating its behavior. In many cases, the user is not interested in implementation details of fine granularity, and thus, one can abstract these details in the presentation of the transducer. Some methods, especially in the context of probabilistic systems, tend to extract the important parts of the implementation by pruning non-relevant behavior according to a notion of importance. An example of such an approach was presented in \cite{Brazdil2015}, where the importance of a state in an implementation is determined by the probability of visiting the state by the strategy.

In an inverse fashion one can incrementally construct the complex implementation starting with a coarse abstraction and gradually refine it with respect to a given partial order, that forces a correct construction. Inspired by counterexample guided abstraction refinement, a series of incremental synthesis procedures have been investigated \cite{7040368, peter+mattmueller/rtss/09, 7519063}. In each stage, refinement suggestions give information on what behavior is added or excluded from the implementation. Allowing to observe each refinement step gives a clearer picture regarding the behavior of the implementation. 
 
In all the approaches above, the product of the synthesis procedure is a representation of a transducer. Although transducers are easy to process, they are not necessarily adequate for presenting the synthesis result to the user. The main reason for that is, that in many domains a transducer is not a standard model users tend to work with in their daily projects. 
Developers of cyber-physical systems for example are familiar with dataflow models. Approaches adapting the idea of synthesizing dataflow models compatible with Simulink\footnote{https://de.mathworks.com/products/simulink.html} or SCADE have become a target of investigation\footnote{http://www.ansys.com/Products/Embedded-Software/ANSYS-SCADE-Suite.}.
Instead of directly synthesizing a transducer as in standard LTL synthesis, an actor-based controller using a computational model of synchronous dataflow (SDF) is considered \cite{cheng2016structural,cheng2017autocode4}. An actor-based controller defines input and output ports and a set of actors and their wiring. The advantage of actor-based controller is that they abstract implementations details that might not be necessary at first for understanding the behavior of the controller. 

\section{Conclusion}
In this paper, we discussed a number of challenges in automatic synthesis of reactive systems. We presented a list of errors that may happen during the specification process and tools for handling the unrealizability and incompleteness of specification, such as identifying unrealizable cores and vacuous parts of the specification.
We also described what obstacles one encounters when trying to understand the outcome of the synthesis process. We explored  different artifacts that can be generated to debug specifications and to 
reason about implementations. Finally, we described different representations for implementations; depending on the domain expertise of the specification designers, synthesis tools should consider which representation would be most beneficial for their target users.

This paper should be seen as an initiator for a  broad discussion on how far synthesis has come and how to make it more attractive for users. Also what further tools are needed to aid the user in the specification process and how to make the outcome of the synthesis process more readable and understandable.

\bibliographystyle{eptcs}
\bibliography{generic}
\end{document}